\documentclass[fleqn,12pt,twoside]{article}
\usepackage{espcrc1}

\usepackage{epsfig}
\usepackage[figuresright]{rotating}

\newcommand{\AmS}{{\protect\the\textfont2
  A\kern-.1667em\lower.5ex\hbox{M}\kern-.125emS}}

\title{Is there elliptic flow without transverse flow?}

\author{P. Huovinen\address{Lawrence Berkeley National Laboratory,
                            Berkeley, USA}\thanks{P.H. acknowledges financial
                          support from the DOE under Contract No. 
                          DE-AC03-76SF00098.},
        P.F. Kolb\address{Institut f\"ur Theoretische Physik,
                          Universit\"at Regensburg, Regensburg, 
                          Germany}$^{\mbox{\scriptsize ,c}}$
        and
        U. Heinz\address{Department of Physics, The Ohio State
                              University, Columbus, USA}}
       
\begin{document}

% typeset front matter
\maketitle

\begin{abstract}
Azimuthal anisotropy of final particle distributions was originally
introduced as a signature of transverse collective flow~\cite{Ollitrault}. 
We show that finite anisotropy in momentum space can result solely
from the shape of the particle emitting source. However, by
comparing the differential anisotropy to recent data from STAR
collaboration~\cite{STAR,Raimond} we can exclude such a scenario,
but instead show that the data favour strong flow as resulting
from a hydrodynamical evolution~\cite{letter1,letter2}.
\end{abstract}
\section{Bjorken Brick}

The intensity of particles emitted from a steady thermal source depends on
the size of the source facing the direction of emitted particles. If the
shape of such a source is not azimuthally symmetric, the particle yield is
anisotropic too. Thus it is easy to parametrize a source with an essentially
arbitrary value of the elliptic flow coefficient $v_2$, but the transverse
momentum dependence of the differential $v_2(p_t)$ is much less trivial.

\begin{figure}[htb]
\begin{minipage}[t]{80mm}
  \begin{center}
    \epsfysize 40mm  \epsfbox{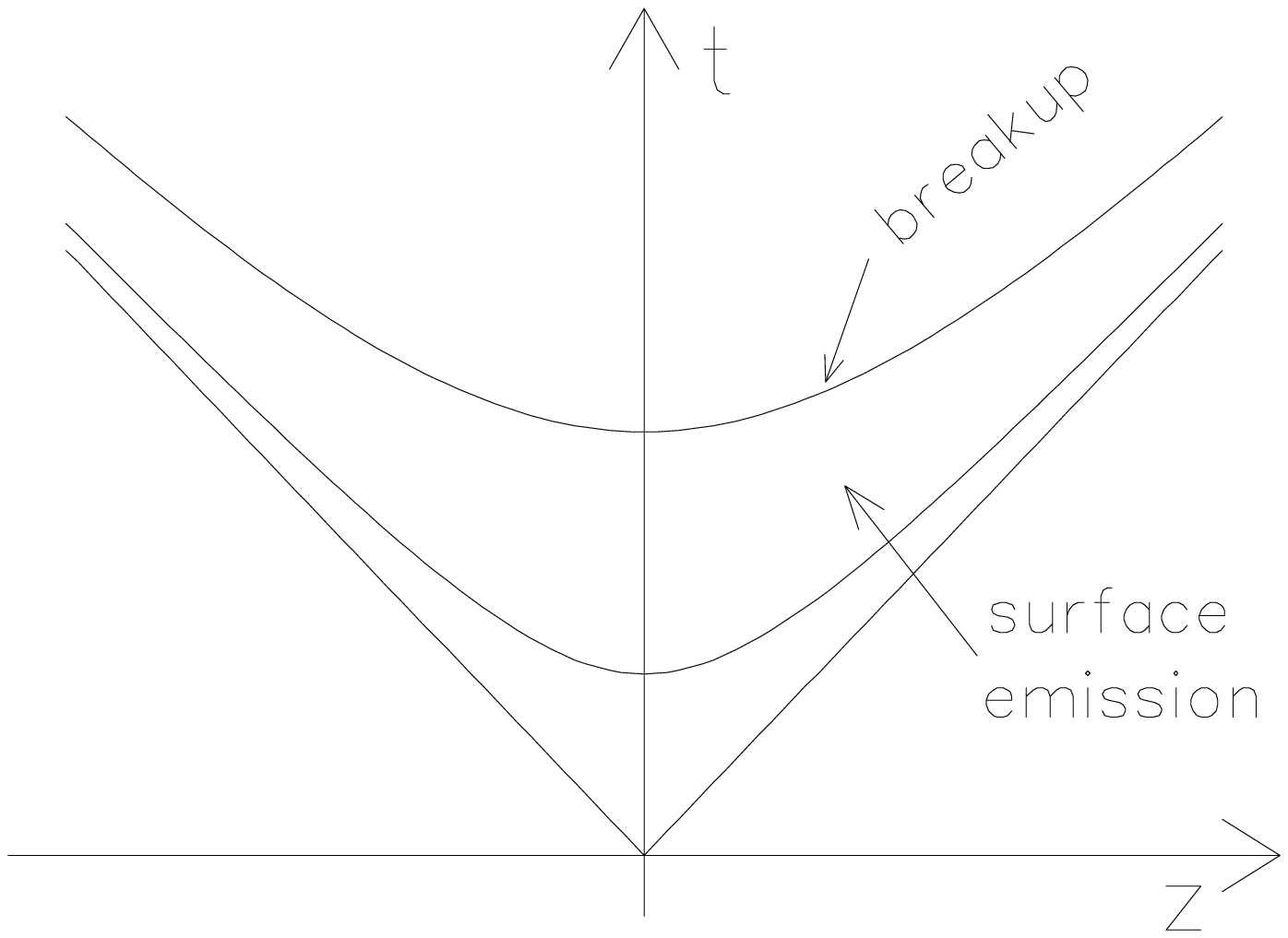}
  \end{center}
\caption{Boost-invariant expansion of the source and emission in two stages.}
\label{spacetime}
\end{minipage}
\hspace{\fill}
\begin{minipage}[t]{75mm}
  \begin{center}
    \epsfysize 40mm \epsfbox{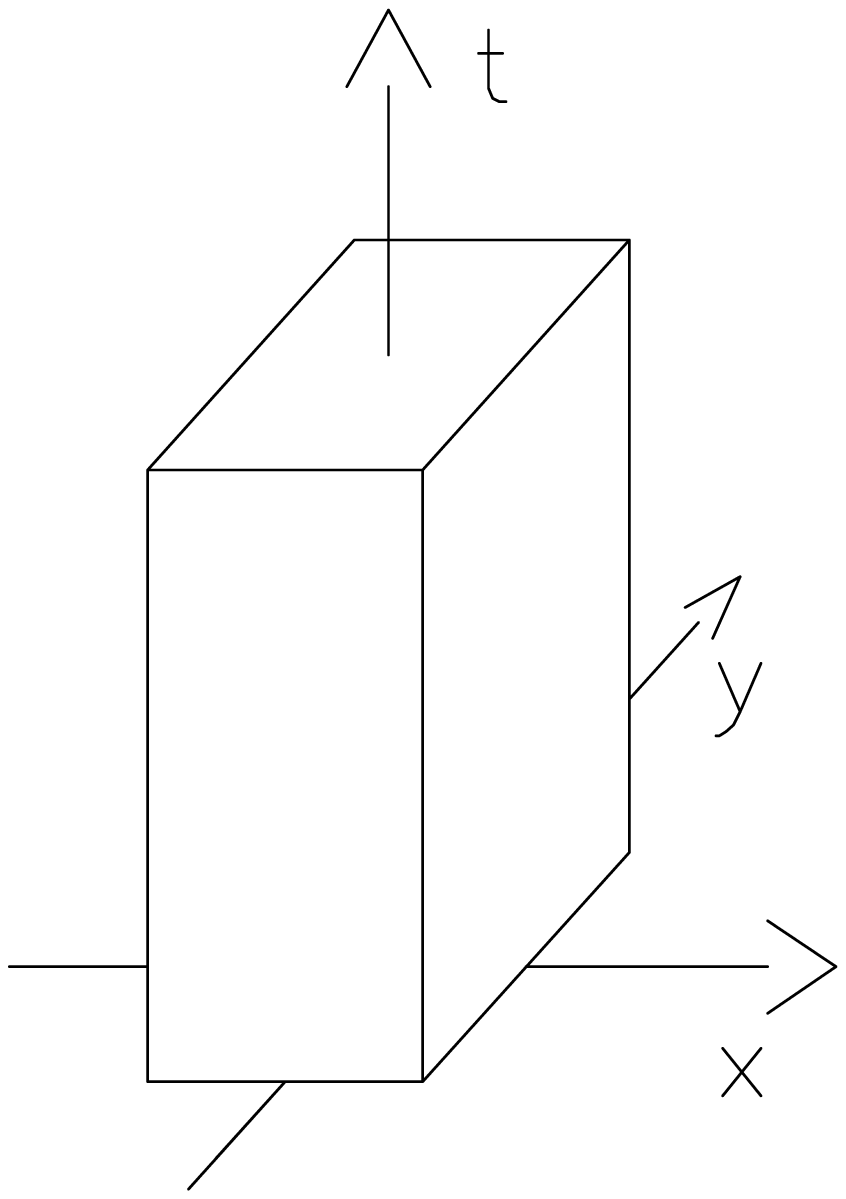}
  \end{center}
\caption{Projection of the Bjorken brick to $xyt$-space.}
\label{brick}
\end{minipage}
\end{figure}

To gain insight what kind of elliptic flow might result from surface effects
only, we constructed a simple parametrization of the source. Let us assume
that the longitudinal expansion of the source is boost invariant, but that
the source does not expand transversally at all. To simplify the calculation
we take the source to be rectangular in transverse plane and to have a
constant temperature $T$. Particle emission now takes place on the surfaces
of the system during its lifetime (from thermalization time $\tau_0$ until
time $t$) and from its entire volume at final breakup at time $t$ (see
Figs.~\ref{spacetime} and~\ref{brick}). Unlike
in~\cite{Urs}, where the source was parametrized using a Gaussian profile,
our source contains emission also through surfaces which normal vectors are
spacelike. If these surfaces are opaque, this allows us to obtain finite
$v_2$ even if transverse flow is absent.

We calculate particle distributions using
%Particle emission from such a source can be calculated using 
a modified Cooper-Frye procedure~\cite{Bugaev}:
% as suggested in~\cite{Bugaev}:
\begin{equation}
 E\frac{dN}{d^3p} 
   = \int d \sigma_\mu p^\mu f(p\cdot u,T)\, \Theta (d \sigma_\mu p^\mu),
 \label{CF}
\end{equation}
where the $\Theta$-function is needed to remove negative contributions to
particle number. For the freeze-out surface described above, the integral 
can be calculated analytically. Using the Boltzmann approximation and the
customary definition of the coefficient $v_2$, we get
\begin{equation}
  v_2(p_t)
   = \frac{(t^2-\tau_0^2)\, p_t K_0\!\left( \frac{m_t}{T} \right) (l_y-l_x)}
          {3 \left(\pi\,t\,m_t K_1\!\left( \frac{m_t}{T} \right) l_xl_y
      + (t^2-\tau_0^2) p_t K_0\!\left( \frac{m_t}{T} \right) (l_y+l_x)\right)}.
  \label{v2}
\end{equation} 

Differential elliptic flow is now essentially controlled by three parameters: 
the ratio of the system size in $x$- and $y$-directions $l_x/l_y$, the ratio
of lifetime to spatial dimension $t/l_y$ and temperature $T$. The effect of
thermalization time $\tau_0$ can be compensated by changing the lifetime and
therefore we prefer to keep it fixed to value $\tau_0 = 0.6$ fm/$c$ inspired
by our hydrodynamical calculations~\cite{letter1}.  

\begin{figure}[htb]
\begin{minipage}[t]{80mm}
  \begin{center}
    \epsfxsize 65mm  \epsfbox{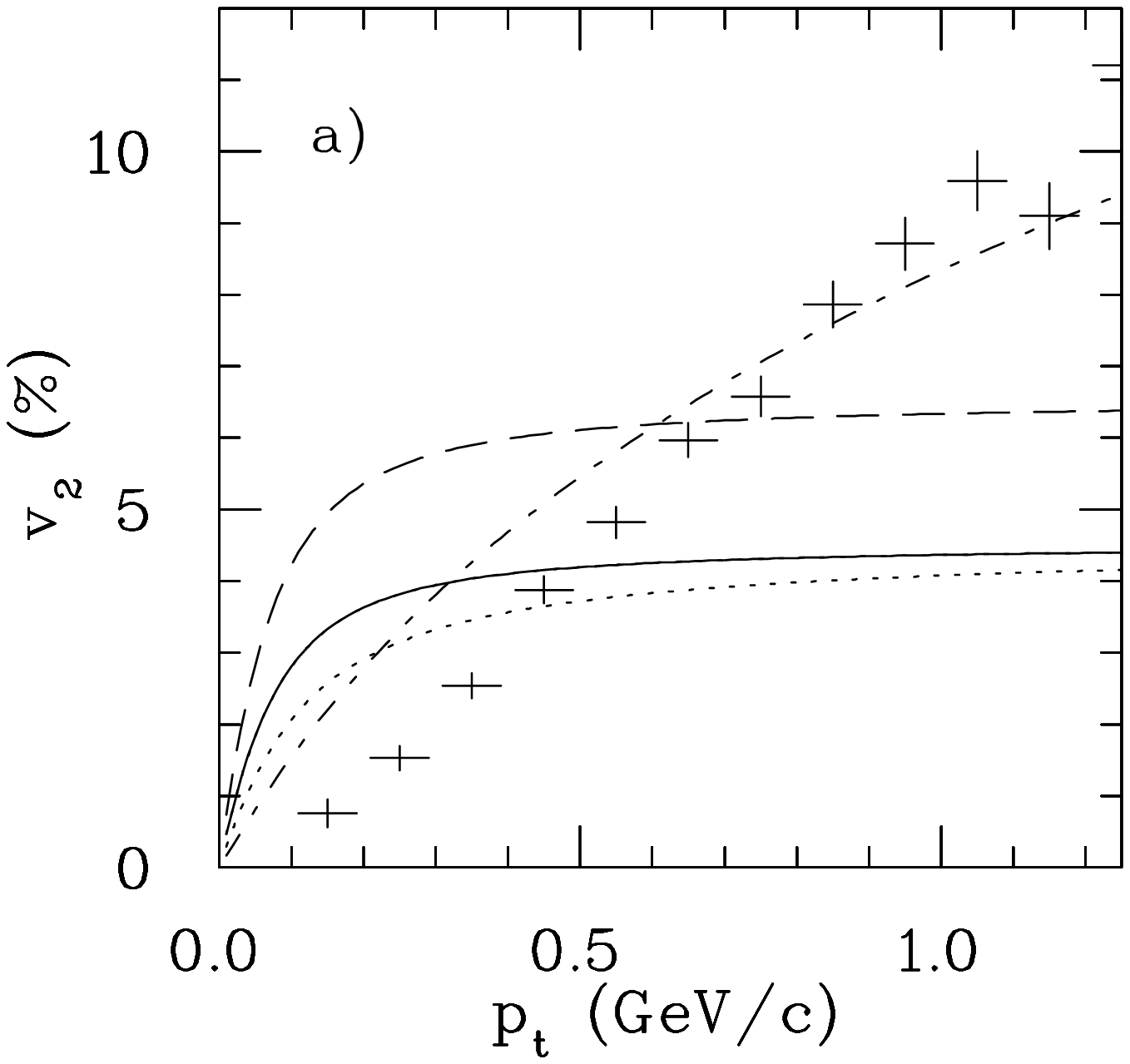}
  \end{center}
\end{minipage}
\hspace{\fill}
\begin{minipage}[t]{75mm}
  \begin{center}
    \epsfxsize 65mm \epsfbox{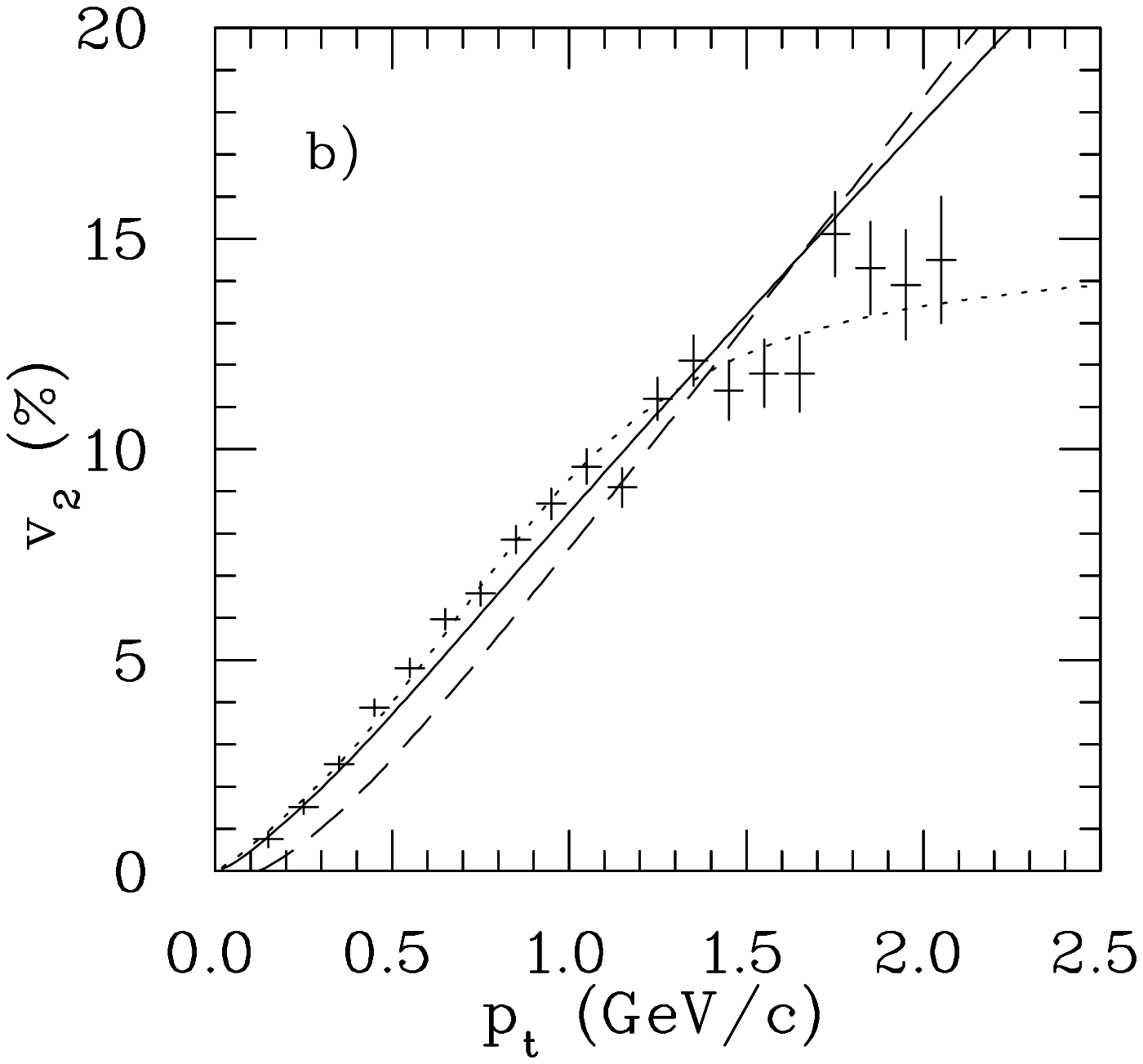}
  \end{center}
\end{minipage}
\caption{Differential elliptic flow of pions from Bjorken brick without flow
         (left) or with additional flow (right) compared to elliptic flow of
         charged particles at RHIC~\cite{STAR}. For explanation of the curves,
         see the text.}
\label{brickresults}
\end{figure}

In Fig.~\ref{brickresults}\,a) the differential elliptic flow of pions
obtained from this model using four different parameter sets is shown.
The solid line corresponds to a case where parameter values are motivated
by the values from hydrodynamical simulations of semi-central collisions
($l_x/l_y=0.58$, $t/l_y = 1.35$, $T=140$ MeV). The qualitative behaviour
of $v_2$ is surprisingly similar to that seen in the Low Density Limit
approximation~\cite{letter1} where the anisotropy is estimated by
calculating a first order correction to the free streaming distribution.
At low values of transverse momenta, $v_2$ increases rapidly but saturates
already around $p_t = 200$ -- 300 MeV/$c$. As demonstrated by the dashed
line obtained using $l_x/l_y = 0.5$ and $t/l_y = 1$, a change in the
ratios scales the saturation value of $v_2$, but does not affect the
value of $p_t$ where saturation takes place.

The only variable which affects the slope of $v_2(p_t)$ is the temperature
of the system as shown by the dotted line ($T=500$ MeV). The effect is
weak and even outrageous values of parameters ($l_x/l_y=1/15$, T=75 GeV,
dashed-dotted line) fail to reproduce the data. Therefore we may conclude
that the present data cannot be described without the presence of strong
transverse flow at RHIC.

The generalization of the Bjorken brick to allow transverse flow is
straightforward although the resulting expression for $v_2(p_t)$ is much
more complicated than Eq.~(\ref{v2}). In addition to the parameters mentioned
earlier, we have the flow velocities in $x$- and $y$-direction $v_x$ and $v_y$.
This generalized parametrization allows us to check which one contributes
more to the observed anisotropy: anisotropic flow field or source shape.
Figure~\ref{brickresults}\,b) shows our fits to the data assuming that
both source shape and velocity field are anisotropic (solid line, $v_x=0.6$,
$v_y=0.57$, $l_x/l_y=0.93$), only the velocity field is anisotropic (dashed
line, $v_x=0.6$, $v_y=0.57$, $l_x/l_y=1$) or only source shape is anisotropic
(dotted line, $v_x=v_y=0.7$, $l_x/l_y=0.73$). Due to the crudeness of the model
the differences between these fits are not meaningful, but the data allows all
three different possibilities. However, an isotropic velocity field from a
deformed source depicts very interesting behaviour of saturating $v_2$ around
$p_t = 1.5$ -- 2 GeV/$c$, in the same region where data do. Whether this is
anything but a coincidence and whether it is possible to produce this kind of
behaviour using a dynamical model remains to be seen.

\section{Hydrodynamic calculations}

We have used a boost invariant hydrodynamical model~\cite{Peter}
to describe the time evolution of the collision system at
RHIC~\cite{letter1,letter2,letter3}. In our approach the initial
conditions of the hydrodynamical evolution were fixed by scaling the
initial densities used to reproduce SPS data~\cite{Peter} until the final
particle multiplicity measured by PHOBOS collaboration~\cite{Phobos} was
reached. In the most central collisions this leads to maximum energy
densities of $\epsilon = 22.3$ -- 23 GeV/fm$^3$ and temperatures of
$T = 270$ -- 330 MeV, depending on the equation of state (EoS).

\begin{figure}[htb]
\begin{minipage}[t]{75mm}
  \begin{center}
    \epsfxsize 65mm  \epsfbox{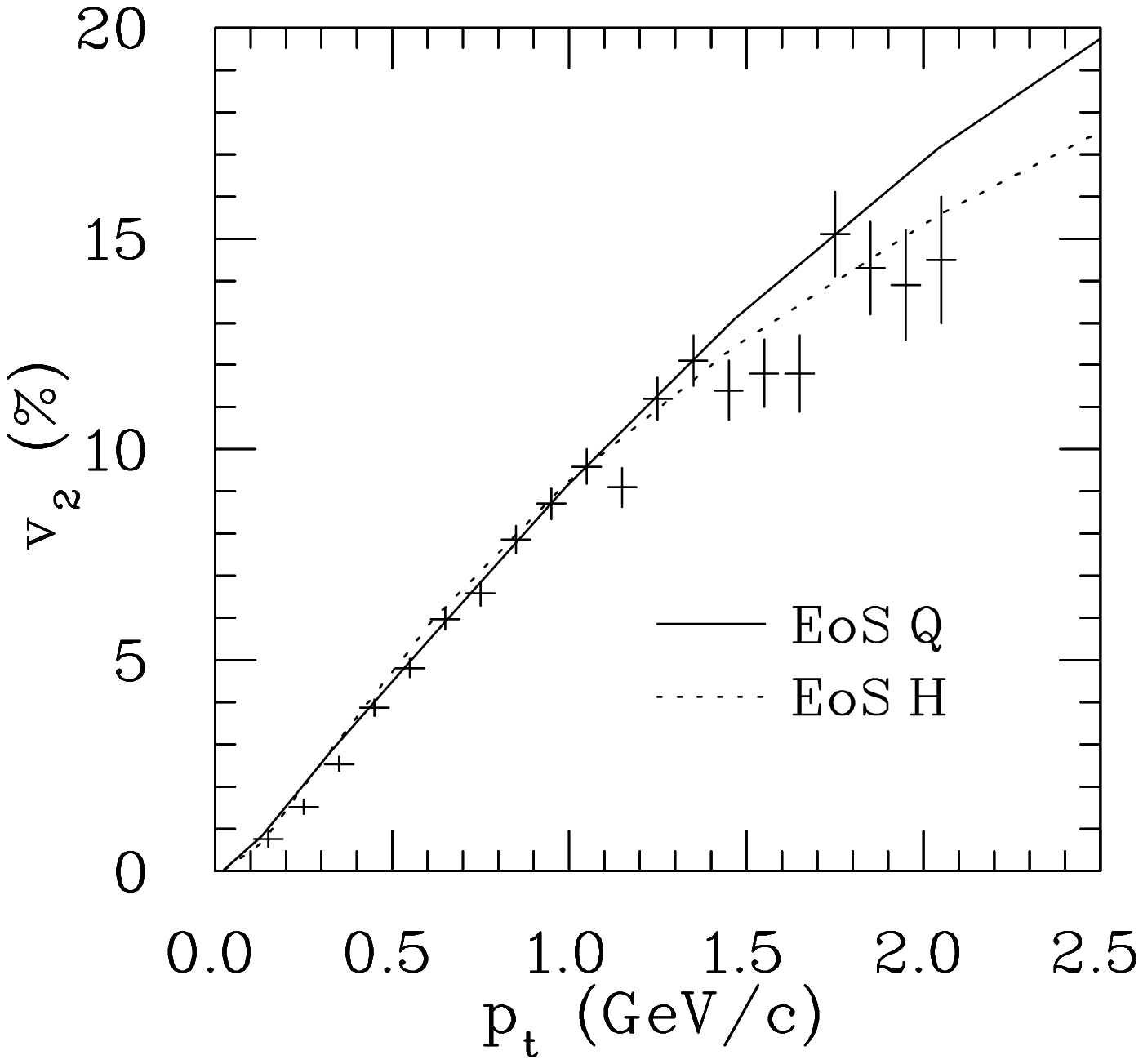}
  \end{center}
\caption{Differential elliptic flow of charged particles in minimum bias
         collisions at RHIC compared to data~\cite{STAR}.}
\label{hydroch}
\end{minipage}
\hspace{\fill}
\begin{minipage}[t]{80mm}
  \begin{center}
    \epsfxsize 76mm \epsfbox{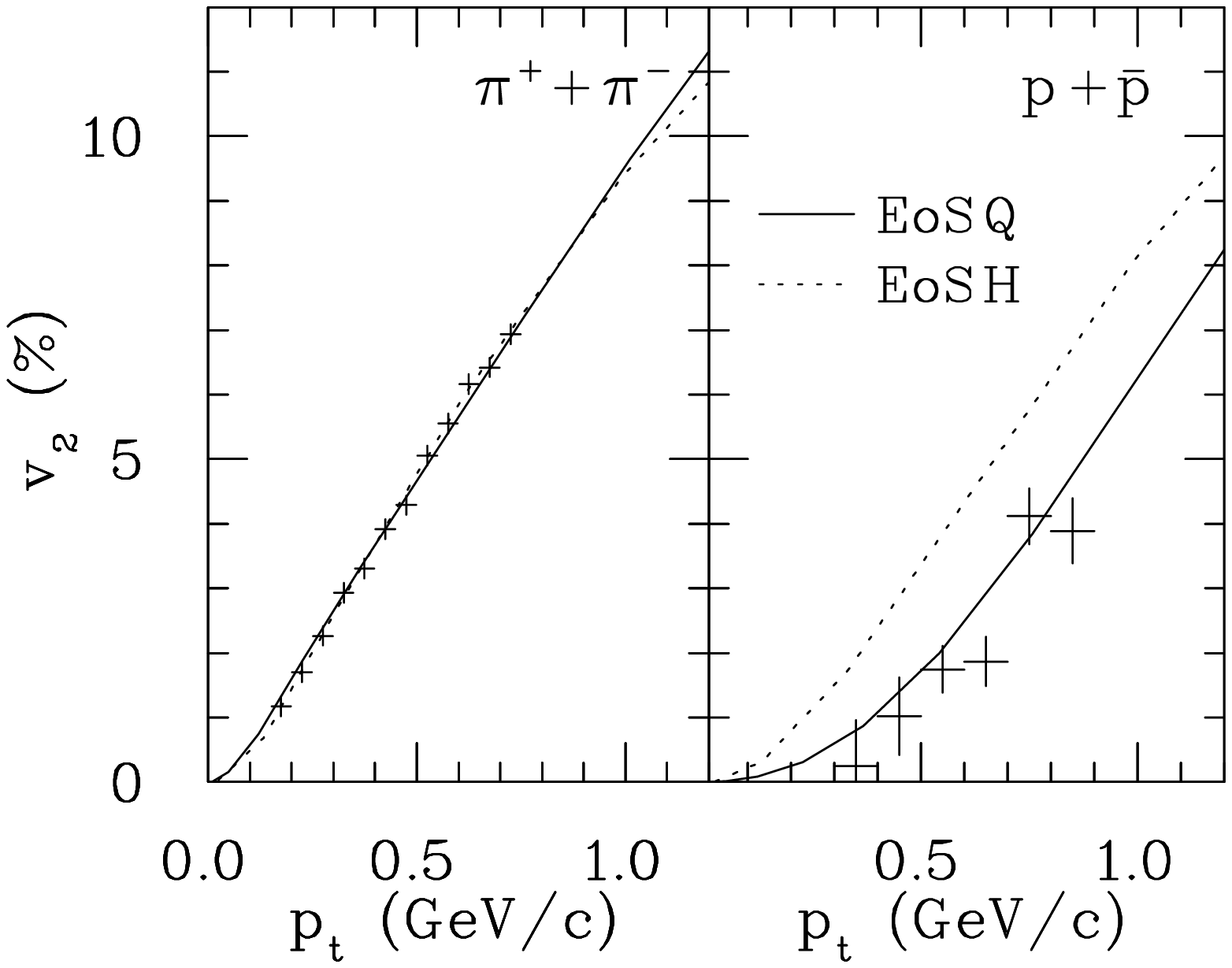}
  \end{center}
\caption{Differential elliptic flow of pions and protons+antiprotons
         in minimum bias collisions at RHIC compared to
         data~\cite{Raimond}.}
\label{hydropipr}
\end{minipage}
\end{figure}

In Figs.~\ref{hydroch} and~\ref{hydropipr} we show our results for minimum
bias collisions obtained using two different EoSs --- one with a phase
transition to QGP (EoS Q) and one without (EoS~H). The agreement between
the charged particle data and hydrodynamic calculation is excellent up to
$p_t\approx 1.5$ GeV/$c$ where the measured elliptic flow begins to show
signs of saturation to some value whereas the hydrodynamic result keeps
increasing. The pion data show similar agreement independent of the EoS
while at first glance the proton $v_2(p_t)$ data seem to favour the
existence of a phase transition. However, as shown in~\cite{letter2,letter3},
proton elliptic flow is sensitive to freeze-out temperature and details of
initial distributions, contrary to the elliptic flow of pions and charged
particles. Therefore no conclusions about the EoS can be drawn without
further constraints on the model. The data also show a strong mass
dependence of $v_2(p_t)$ which is typical to hydrodynamical models:
the heavier the particle, the smaller $v_2$ at small values of $p_t$.

As a function of centrality (see~\cite{Raimond,letter1}) our fit to data
is less successful. As can be expected, we can reproduce the data in central
and semi-central collisions, but we overestimate elliptic flow in peripheral
collisions, due to the loss of a sufficient degree of thermalization. 

The observed elliptic flow depends on the amount of rescatterings during
the expansion. Hydrodynamics assumes zero mean free path, or equivalently,
infinitely strong rescattering. It therefore provides practical upper limits
for the anisotropy of particle distribution. That the experimental data reach
these limiting values for elliptic flow and also reflect the mass dependence
typical for a hydrodynamic system points towards early thermalization and
a subsequent hydrodynamic evolution in central and semi-central %Au+Au
collisions at RHIC.

\end{document}